# Artificial non-Abelian lattice gauge fields for photons in the synthetic frequency dimension

Dali Cheng,[1] Kai Wang,[1] and Shanhui Fan[1,*]

[1]*Ginzton Laboratory and Department of Electrical Engineering, Stanford University, Stanford, California 94305, USA*

*shanhui@stanford.edu

**Abstract**

Non-Abelian gauge fields give rise to nontrivial topological physics. Here we develop a scheme to create an arbitrary SU(2) lattice gauge field for photons in the synthetic frequency dimension using an array of dynamically modulated ring resonators. The photon polarization is taken as the spin basis to implement the matrix-valued gauge fields. Using a non-Abelian generalization of the Harper–Hofstadter Hamiltonian as a specific example, we show that the measurement of the steady-state photon amplitudes inside the resonators can reveal the band structures of the Hamiltonian, which show signatures of the underlying non-Abelian gauge field. These results provide opportunities to explore novel topological phenomena associated with non-Abelian lattice gauge fields in photonic systems.

*Introduction*—An important theme of physics research in the past decades is to use gauge fields to create systems with nontrivial topological properties. A canonical example is the quantum Hall effect [1], where a U(1) gauge field is coupled to electrons in two dimensions, resulting in electronic band structures with a nontrivial topology and topologically robust edge states. As a generalization of the quantum Hall effect, a topological insulator can be described as electrons coupled to a spin-dependent Abelian gauge field [2,3]. More recently, there is emerging interest in more complex topological physics based on non-Abelian gauge fields [4-14]. While the concept of gauge fields was initially developed for electronic systems, in recent years there are significant efforts in synthesizing gauge fields for other systems including photons [15-23], phonons [24-30] and cold atoms [7,31-43].

Different from the U(1) gauge field which manifests in nature as the magnetic field coupled to electrons, non-Abelian gauge fields do not naturally exist for either electrons or photons. Therefore, to explore the associated physics, non-Abelian gauge fields need to be artificially synthesized [44], as has been demonstrated on diverse experimental platforms, including cold atoms [7,31-34,36,40,42,43], exciton-polaritons [45-51], mechanics and acoustics [24,28], superconducting circuits [52], electrical circuits [53], and molecular systems [54]. For photons, there have also been recent works that create non-Abelian gauge fields [15,16,21,22]. These works, however, have not incorporated the synthetic non-Abelian gauge fields into lattice systems to achieve topological insulator physics.

In this Letter we propose to create a lattice system with an arbitrary SU(2) non-Abelian gauge field for photons in the synthetic frequency dimension. The concept of synthetic dimensions is to couple internal degrees of freedom of particles to form a synthetic lattice system [55-80]. In a photonic ring resonator, for example, electro-optic modulators can be incorporated to couple different longitudinal modes together to form a synthetic frequency lattice [66]. The Hamiltonian of such a synthetic frequency lattice can be programmed by designing the waveforms of the electrical signals that control the modulators, which results in significant tunability and flexibility in creating different Hamiltonians [63,66]. Artificial U(1) gauge fields have been achieved for photons in synthetic frequency dimensions [18,20].

To create an arbitrary SU(2) non-Abelian lattice gauge field in the synthetic frequency dimension, we consider a system consisting of an array of coupled ring resonators. We take the polarization as the pseudo-spin for photons, and each resonator incorporates two electro-optic modulators and two polarization rotators. We show that a non-Abelian gauge field in a lattice can be achieved with an appropriate configuration of these polarization rotators and modulators. As an example, we implement an SU(2) generalization of the Harper–Hofstadter Hamiltonian, the U(1) version of which is directly related to the quantum Hall effect. Using numerical calculations, we show that by measuring the steady-state photon amplitudes inside the resonators, the bulk states, topologically protected edge states and the non-Abelian 'Hofstadter butterflies' can be directly probed, all of which show signatures of the underlying non-Abelian gauge field. The results here point to a pathway to artificially synthesize non-Abelian lattice gauge fields for photons and to explore the associated topological physics.

*Results*—Figure 1 illustrates our setup to create the non-Abelian gauge field in a two-dimensional lattice. As shown in Fig. 1(b), Our setup consists of a one-dimensional array of coupled ring resonators. For each resonator, the main section of the waveguides supports two electromagnetic modes with orthogonal polarizations, denoted as the $s$- and $p$-polarizations. We use these two polarizations to form the pseudo-spin state basis for the photons, with the $s$ and $p$ polarizations corresponding to the spin up and down states, respectively. We assume that the two polarizations are degenerate in the main section, and the group velocity dispersion of the waveguides is negligible. Hence the resonator has the same free spectral range $\Omega_R$ for both polarizations. Neighboring resonators are coupled to each other. Because of the nature of coupling, light in the neighboring resonators propagates in opposite clockwise or counter-clockwise directions. We assume that the coupling preserves the photon spin state, and the coupling strength $\kappa$ is real, frequency-independent, and spin-independent.

Within each resonator [Fig. 1(c)], we incorporate two polarization rotators in the main section of the waveguide. The polarization rotators perform SU(2) transformations on the photon spin state [81]. The main section of the waveguide is connected with a polarization splitter and a combiner, which separate the spin-up and spin-down states into two different polarization-maintaining waveguide branches. Within each branch there is an electro-optic modulator. These two modulators in two branches are independently controlled and modulated at the frequency $\Omega_R$.

The coupled resonator array is modeled by the tight-binding lattice structure shown in Fig. 1(a). The coupling between neighboring resonators forms one spatial dimension, labeled by the resonator index $m$. The frequency modes of the ring resonators form a synthetic dimension, labeled by the mode index $n$. With the modulation frequency of the modulators chosen at the free spectral range $\Omega_R$ of the resonators, the neighboring frequency modes in each resonator are coupled [20]. In the following, we show that by properly choosing the parameters for the polarization rotators and the modulators, an arbitrary SU(2) gauge field can be implemented in the two-dimensional lattice in Fig. 1(a).

We start by examining the $m$-th resonator in the array by itself and show that along the synthetic frequency dimension, this resonator can implement the Hamiltonian

$$\hat{H}_m = -g \sum_n (\hat{a}^\dagger_{m,n+1} U_m \hat{a}_{m,n} + \text{H.c.}) \tag{1}$$

with proper design of the polarization rotators and the modulators. (We assume $\hbar = 1$ throughout the Letter.) In Eq. (1), $g$ is the real coupling strength between neighboring sites along the synthetic frequency dimension. $\hat{a}_{m,n} = [\hat{a}_{m,n,\uparrow}, \hat{a}_{m,n,\downarrow}]^\text{T}$ with $\hat{a}_{m,n,\uparrow}$ and $\hat{a}_{m,n,\downarrow}$ being the annihilation operators for the spin-up and spin-down states on the $(m, n)$ lattice site, respectively. $\hat{a}^\dagger_{m,n} = [\hat{a}^\dagger_{m,n,\uparrow}, \hat{a}^\dagger_{m,n,\downarrow}]$. $U_m$, the link variable along the synthetic frequency dimension, is an element of the SU(2) matrix group, and can be explicitly written as

$$U_m = \begin{bmatrix} e^{i\alpha_m} \cos\theta_m & e^{-i\beta_m} \sin\theta_m \\ -e^{i\beta_m} \sin\theta_m & e^{-i\alpha_m} \cos\theta_m \end{bmatrix}, \quad -\pi < \alpha_m, \beta_m, \theta_m \leq \pi \tag{2}$$

$U_m = \exp(iA_m)$. $A_m$, a 2×2 Hermitian matrix, is the corresponding gauge potential.

To show Eq. (1), we expand the electric field inside the $m$-th resonator at the time $TT_R + t$ as

$$\psi(m, t, T) = \sum_n a_{m,n}(T) e^{in\Omega_R t} \tag{3}$$

Here $\psi(m, t, T) \equiv [\psi_\uparrow(m, t, T), \psi_\downarrow(m, t, T)]^\text{T}$ describes the spin-dependent amplitude of the electric field. $T_R = 2\pi/\Omega_R$ is the round-trip propagation time of light inside the resonator. $t \in (-T_R/2, T_R/2]$ is the fast time variable for light propagation within each round trip, and $T$, a non-negative integer, is the slow time variable indexing the number of round trips [59]. $a_{m,n}(T) \equiv [a_{m,n,\uparrow}(T), a_{m,n,\downarrow}(T)]^\text{T}$ describes the spin-dependent amplitudes on the lattice site labelled by $(m, n)$. After the light passes through a round trip inside the resonator, the electric field becomes

$$\psi(m, t, T+1) = \sum_n a_{m,n}(T+1) e^{in\Omega_R t} = V_m D_m(t) V_m^\dagger \sum_n a_{m,n}(T) e^{in\Omega_R t} \tag{4}$$

where $V_m$ and $V_m^\dagger$ describe the time-independent polarization rotations, and $D_m(t)$ describes the time-dependent, spin-dependent modulation in Fig. 1(c). We take the transmission functions of the modulation as

$$D_m(t) = \begin{bmatrix} e^{i2gT_R \cos(\Omega_R t + \delta_m)} & 0 \\ 0 & e^{i2gT_R \cos(\Omega_R t - \delta_m)} \end{bmatrix} \approx \mathbb{I}_2 + i2gT_R \begin{bmatrix} \cos(\Omega_R t + \delta_m) & 0 \\ 0 & \cos(\Omega_R t - \delta_m) \end{bmatrix}$$
(5)

The modulation frequency is $\Omega_R$, the modulation strength is $2gT_R$, and the two modulators differ in their phases by $2\delta_m$. $\mathbb{I}_N$ is the $N \times N$ identity matrix, and the approximation in Eq. (5) is valid in the weak-modulation limit $gT_R \ll 1$. Inserting Eq. (5) into Eq. (4), we obtain

$$a_{m,n}(T+1) - a_{m,n}(T) = igT_R[V_m e^{i\delta_m \sigma_z} V_m^\dagger a_{m,n-1}(T) + V_m e^{-i\delta_m \sigma_z} V_m^\dagger a_{m,n+1}(T)]$$
(6)

where $\sigma_x$, $\sigma_y$ and $\sigma_z$ are the Pauli matrices for the photon spin. By defining $\partial_T a_{m,n} \approx [a_{m,n}(T+1) - a_{m,n}(T)]/T_R$ [59], we obtain

$$i\partial_T a_{m,n} = -g(V_m e^{i\delta_m \sigma_z} V_m^\dagger a_{m,n-1} + V_m e^{-i\delta_m \sigma_z} V_m^\dagger a_{m,n+1})$$
(7)

Comparing Eq. (7) and the Hamiltonian in Eq. (1), we find that they describe the same dynamics if $U_m = V_m e^{-i\delta_m \sigma_z} V_m^\dagger$. Therefore, to implement the link variable $U_m$ described by Eq. (2), the parameters for the modulators and the polarization rotators are chosen as

$$\delta_m = \cos^{-1}(\cos\theta_m \cos\alpha_m)$$
$$V_m = e^{-i\left(\frac{\beta_m}{2} + \frac{\pi}{4}\right)\sigma_z} e^{i\gamma_m \sigma_y}, \quad \gamma_m = \frac{1}{2}\text{Arg}(-\cos\theta_m \sin\alpha_m + i\sin\theta_m)$$
(8)

where Arg() is the principal value of the complex argument in the range $(-\pi, \pi]$.

Next, we take into account the coupling between neighboring resonators in the array. In the weak-coupling and weak-modulation regime, i.e., $\kappa T_R \ll 1$ and $gT_R \ll 1$, the Hamiltonian describing the entire resonator array in Fig. 1(b) is

$$\hat{H} = -\sum_{m,n}\left(\kappa \hat{a}_{m+1,n}^\dagger \hat{a}_{m,n} + g \hat{a}_{m,n+1}^\dagger U_m \hat{a}_{m,n} + \text{H.c.}\right)$$
(9)

The loop operator of each square plaquette in this lattice in the counter-clockwise direction, with lattice site $(m, n)$ being the left-bottom corner, reads

$$W_{m,n} = \kappa^2 g^2 U_{m+1} U_m^\dagger$$
(10)

In the special case where $U_m = e^{im\phi}\mathbb{I}_2$, $W_{m,n} \propto e^{i\phi}\mathbb{I}_2$. The Hamiltonian is then reduced to the standard Harper–Hofstadter model with a U(1) Abelian gauge field, and $\phi$ is the magnetic flux inside each plaquette of the lattice. More generally, the $U_m$'s can be chosen such that $[W_{m,n}, W_{m',n'}] \neq 0$ for $m' \neq$ $m, n' \neq n$. This non-commutativity of the loop operators is the condition for the presence of a non-Abelian gauge field [41].

Here we briefly comment on the values of some parameters in our model. In a 20-meter-long fiber loop cavity with $\Omega_R = 2\pi \times 10$ MHz [70,79,82,83], for example, the round-trip time is $T_R = 0.1$ μs. If we choose $gT_R = 0.05$, which is in the regime of weak modulation, the maximum voltage applied on the electro-optic modulator is approximately $0.03V_\pi$ where $V_\pi$ is the half-wave voltage. The choice of $\kappa T_R = 0.05$, which is in the regime of weak coupling, corresponds to a beam splitter with 90:10 coupling ratio.

We next show that the band structures of a lattice system in the synthetic dimension that possesses non-Abelian gauge fields can be observed by measuring the steady-state photon amplitudes inside the resonators. We assume that there are $M$ coupled resonators in Fig. 1(b). Resonators 1 and $M$ are either coupled or decoupled to implement the periodic or open boundary conditions along the spatial dimension, respectively. To probe the system, we can use an input waveguide to excite one of the resonators [for example, resonator 1 in Fig. 1(b)] by a continuous wave laser with a continuously tunable frequency $\omega_{\text{CW}} = \omega_0 + n\Omega_R + \delta\omega$. $\omega_0$ is the central frequency, and $\delta\omega$ is the frequency detuning from the $n$-th resonance of the cavities. The time-dependent intra-cavity light of each resonator can be measured from a drop port that samples the steady-state photon amplitudes inside the resonator. The two polarization components of the sample can be separated with a polarization beam splitter. The frequency contents of both polarization components can then be obtained using a heterodyne detection scheme where each of these polarization components is interfered with a portion of the frequency-shifted laser input [70]. This measurement provides a detection of $\psi(m, t, \delta\omega) \equiv [\psi_\uparrow(m, t, \delta\omega), \psi_\downarrow(m, t, \delta\omega)]^T$, which is related to $\psi(m, t, T)$ through a Fourier transformation on the variable $T$.

It has been noted that the fast time variable $t$ is directly related to the wavevector $k_f$ associated with the synthetic frequency dimension by $k_f = \Omega_R t$ [82,84-86]. Therefore, below, we will relabel the fast time variable $t$ as $k_f$ so that the connection to band structure becomes more transparent. Assuming that we are exciting the $m_0$-th resonator with a spin state $[\psi_\uparrow^{\text{exc}}, \psi_\downarrow^{\text{exc}}]^T$, based on the input–output formalism [82,87,88], the steady-state spin-dependent amplitudes in the $m$-th resonator, shown here for the spin-up case as an example, are related to the input $|\psi_{\text{in}}\rangle$ by

$$\psi_\uparrow(m, k_f, \delta\omega) = \left\langle m, \uparrow \left| \frac{i\sqrt{\gamma}}{\delta\omega - \hat{H}(k_f) + i\left(\gamma + \frac{\gamma_0}{2}\right)} \right| \psi_{\text{in}} \right\rangle$$

(11)

where $\gamma_0$ is the intrinsic loss of the resonators, and $\gamma$ is the coupling constant between the resonators and the input–output waveguides. $|m, \uparrow\rangle \equiv \left[0, 0, \ldots, \underbrace{1, 0}_{m\text{-th resonator}}, \ldots, 0, 0\right]^T$ and $|\psi_{\text{in}}\rangle \equiv$

$$\left[0,0,\ldots,\underbrace{\psi_\uparrow^{\text{exc}},\psi_\downarrow^{\text{exc}}}_{m_0\text{-th resonator}},\ldots,0,0\right]^T$$ are $2M$-element column vectors describing the spin-dependent amplitudes inside the resonator array.

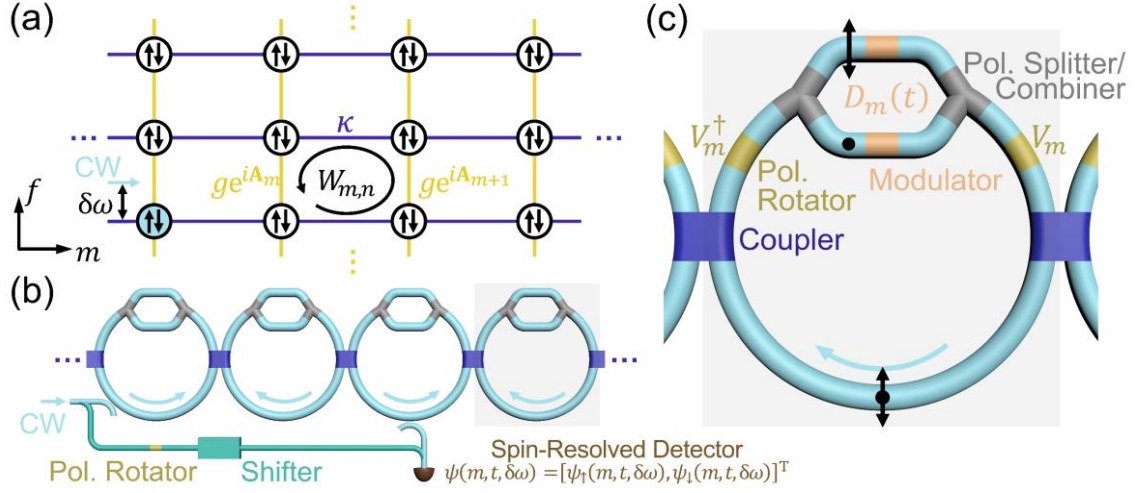

**Fig. 1** Implementation of an arbitrary SU(2) lattice gauge field in the synthetic frequency dimension. (a) The tight-binding lattice model involving the non-Abelian gauge field. Each circle represents a lattice site with spin-up and spin-down states. The lattice site excited by the continuous wave (CW) laser input is colored in blue. (b) A schematic of the coupled resonator array. In the heterodyne detection, a polarization rotator is inserted to ensure that both spin components of the reference signal are finite when mixing. (c) The detailed structure of one resonator as shaded in grey in (b). The black arrows and dots indicate two orthogonal polarizations (spins).

As a concrete example, we study an SU(2) generalization of the Harper–Hofstadter Hamiltonian under Landau gauge. The Hamiltonian is created by setting $U_m = e^{im\theta\sigma_y}e^{im\varphi\sigma_z}$ in Eq. (9), which results in

$$\widehat{H}_1 = -\sum_{m,n}\left(\kappa\hat{a}^\dagger_{m+1,n}\hat{a}_{m,n} + g\hat{a}^\dagger_{m,n+1}e^{im\theta\sigma_y}e^{im\varphi\sigma_z}\hat{a}_{m,n} + \text{H.c.}\right)$$

(12)

With a choice of $\theta/2\pi = p_\theta/q_\theta$, $\varphi/2\pi = p_\varphi/q_\varphi$ where $p_\theta, q_\theta, p_\varphi, q_\varphi$ are integers, the lattice has the period of $q$ along the spatial dimension, where $q$ is the least common multiple of $q_\theta$ and $q_\varphi$. This Hamiltonian has been shown to exhibit rich topological insulator behaviors with spin–orbit-coupled Hofstadter butterfly pairs [9]. Using Eq. (11), we can numerically simulate the measurement of the band structures of the Hamiltonian $\widehat{H}_1$. We systematically study $\widehat{H}_1$ when $q = 3$. It has been shown that $\widehat{H}_1$ is invariant with respect to the interchange of $\theta$ and $\varphi$ [9]. Moreover, we note that the band structure does not change under $\theta \to -\theta$ transformation: $\widehat{H}_1(-\theta,\varphi) = \Sigma_z\widehat{H}_1(\theta,\varphi)\Sigma_z^{-1}$ where $\Sigma_z = \sigma_z \otimes \mathbb{I}_M$.

Therefore, for $q = 3$, to create a spatially varying gauge field, only two inequivalent parameter combinations are available: $(\theta, \varphi) = (0, 2\pi/3)$ or $(2\pi/3, 2\pi/3)$. According to the non-Abelian condition defined through the non-commutativity of the loop operators [9,41], $(\theta, \varphi) = (0, 2\pi/3)$ corresponds to an Abelian gauge field related to the quantum spin Hall effect [2,3], and $(\theta, \varphi) = (2\pi/3, 2\pi/3)$ corresponds to a non-Abelian gauge field. We refer to these two parameter combinations as the 'Abelian case' and the 'non-Abelian case' in the following.

Figure 2 shows the band structures of $\hat{H}_1$ when the periodic boundary condition is applied to the spatial dimension. In Figs. 2(a) and 2(e), we plot the band structures by diagonalizing the Hamiltonian $\hat{H}_1$ for the Abelian and non-Abelian cases, respectively. $k_m$ is the wavevector associated with the spatial dimension $m$. To connect the band structures with the simulated steady-state spin-dependent amplitudes in Eq. (11), we define the $k_m$-dependent, spin-dependent photon amplitudes via the Fourier transform

$$\psi_s(k_m, k_f, \delta\omega) = \sum_{m=1}^{M} \psi_s(m, k_f, \delta\omega) e^{-ik_m m}, \qquad s = \uparrow, \downarrow$$

(13)

Figs. 2(b)–2(d) and 2(f)–2(h) show the simulated results of $|\psi_\uparrow(k_m, k_f, \delta\omega)|$ at different values of $k_m$. The resonance locations of $|\psi_\uparrow(k_m, k_f, \delta\omega)|$ coincide with the slices of the band structures in Figs. 2(a) and 2(e) [89].

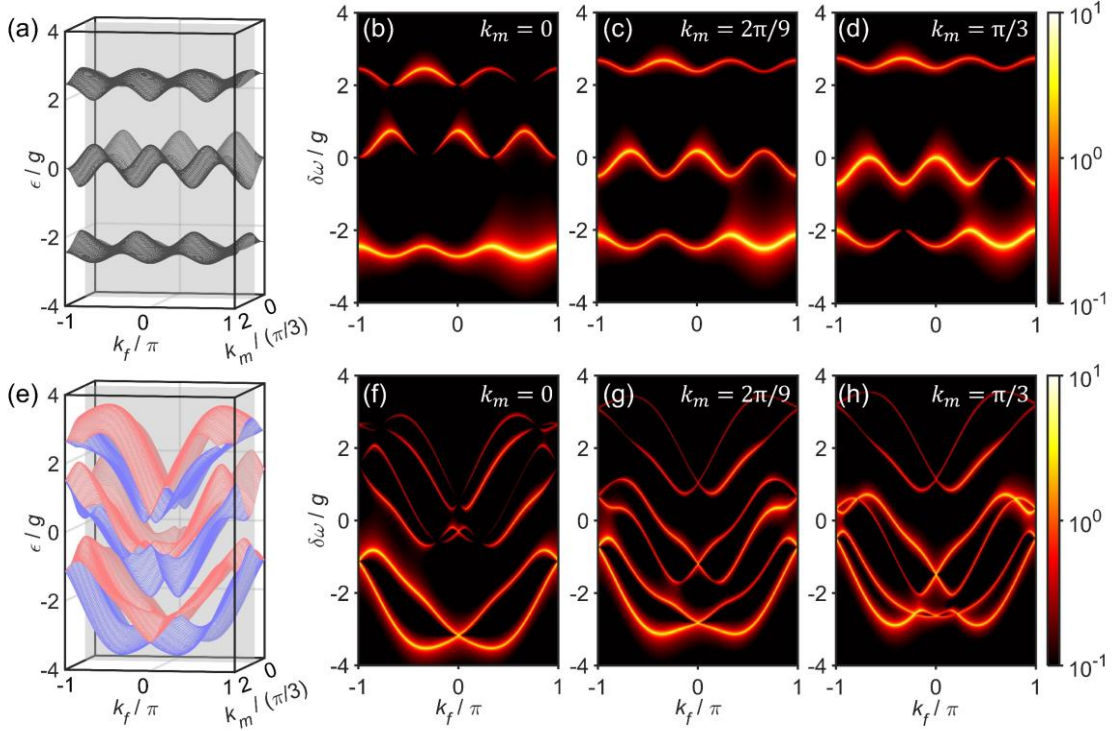

**Fig. 2.** Band structures of the Hamiltonian $\widehat{H}_1$ under the periodic boundary condition. (a)–(d) The Abelian case, $(\theta, \varphi) = (0, 2\pi/3)$. (e)–(h) The non-Abelian case, $(\theta, \varphi) = (2\pi/3, 2\pi/3)$. (a) and (e) are acquired by diagonalizing $\widehat{H}_1$. (b)–(d), (f)–(h) are simulated results of $|\psi_\uparrow(k_m, k_f, \delta\omega)|$ at fixed $k_m$ values, as illustrated by the grey slices in (a) and (e). $\kappa = g$ in this figure, $\gamma = \gamma_0 = 0.01g$, $m_0 = 1$, $\psi_\uparrow^{\text{exc}} = 1$, $\psi_\downarrow^{\text{exc}} = 0$ in (b)–(d), (f)–(h).

In Fig. 3, we truncate the lattice in Fig. 1(a) along the spatial dimension and study the eigenspectra of $\widehat{H}_1$ subject to the open boundary condition, for both the Abelian case [Figs. 3(a)–3(c)] and the non-Abelian case [Figs. 3(d)–3(f)]. In Figs. 3(a) and 3(d), the bulk states of the continuum can be viewed as the bands in Figs. 2(a) and 2(e) projected onto the $k_f$ axis. Within the band gaps, edge states are present. To illustrate the properties of these edge states, as examples in Figs. 3(b) and 3(e), we plot the pseudo-spins $(\langle\sigma_x\rangle, \langle\sigma_y\rangle, \langle\sigma_z\rangle)$ of the edge states on the Bloch sphere, for the energy as indicated by the color dots in Figs. 3(a) and (d) with $m = 1$:

$$\langle\sigma_s\rangle = \langle\psi^{\text{eig}}(m, k_f, \epsilon)|\sigma_s|\psi^{\text{eig}}(m, k_f, \epsilon)\rangle, \qquad |\psi^{\text{eig}}(m, k_f, \epsilon)\rangle = \begin{bmatrix}\psi_\uparrow^{\text{eig}}(m, k_f, \epsilon)\\ \psi_\downarrow^{\text{eig}}(m, k_f, \epsilon)\end{bmatrix}, \qquad s = x, y, z$$

(14)

We use $\psi_s^{\text{eig}}(m, k_f, \epsilon)$ to represent the eigenstate of $\widehat{H}_1$ with wavevector $k_f$, energy $\epsilon$, pseudo-spin $s$ and on lattice site $m$. In the Abelian case, the edge states are purely spin-up when $k_f\epsilon < 0$, and spin-down when $k_f\epsilon > 0$. This is in accordance with the quantum spin Hall effect where two electron spins are decoupled and independently coupled to opposite magnetic fields [2,3,90]. In contrast, in the non-Abelian case, as shown in Figs. 3(d) and 3(e), the pseudo-spins become dependent on $k_f$, but they are still orthogonal for a pair of edge states at $(k_f, \epsilon)$ and $(-k_f, \epsilon)$ as a result of the Kramers' degeneracy [89]. In Figs. 3(c) and 3(f), we perform the steady-state simulations by Eq. (11), and plot the $k_f$-resolved photon number in all cavities:

$$N(k_f, \delta\omega) \equiv \sum_{m=1}^{M} \sum_{s=\uparrow,\downarrow} |\psi_s(m, k_f, \delta\omega)|^2$$

(15)

The excitation for the Abelian case is $\psi_\uparrow^{\text{exc}} = 1$ and $\psi_\downarrow^{\text{exc}} = 0$. For the non-Abelian case, we use a $k_f$-dependent excitation instead:

$$\psi_s^{\text{exc}}(k_f) = \begin{cases} \psi_s^{\text{eig}}(m_0, k_f, \epsilon), & k_f\epsilon < 0 \\ \psi_s^{\text{eig}}(m_0, -k_f, \epsilon), & k_f\epsilon > 0 \end{cases}$$

(16)

where the eigenenergy $\epsilon$ is chosen at the energies of the edge states. In Figs. 3(c) and 3(f), one sees that only the positive-energy (negative-energy) edge state is excited when $k_f < 0$ ($k_f > 0$), which confirms the spin-dependent properties discussed above. In other words, the SU(2) generalization of the Harper–Hofstadter model inherits the chirality of the edge states of the U(1) counterpart, but the pseudo-spin basis becomes $k_f$-dependent because of the spin mixture induced by the non-Abelian gauge field. This unique signature of the non-Abelian gauge field can be probed in our measurement scheme by an appropriate time-dependent input state.

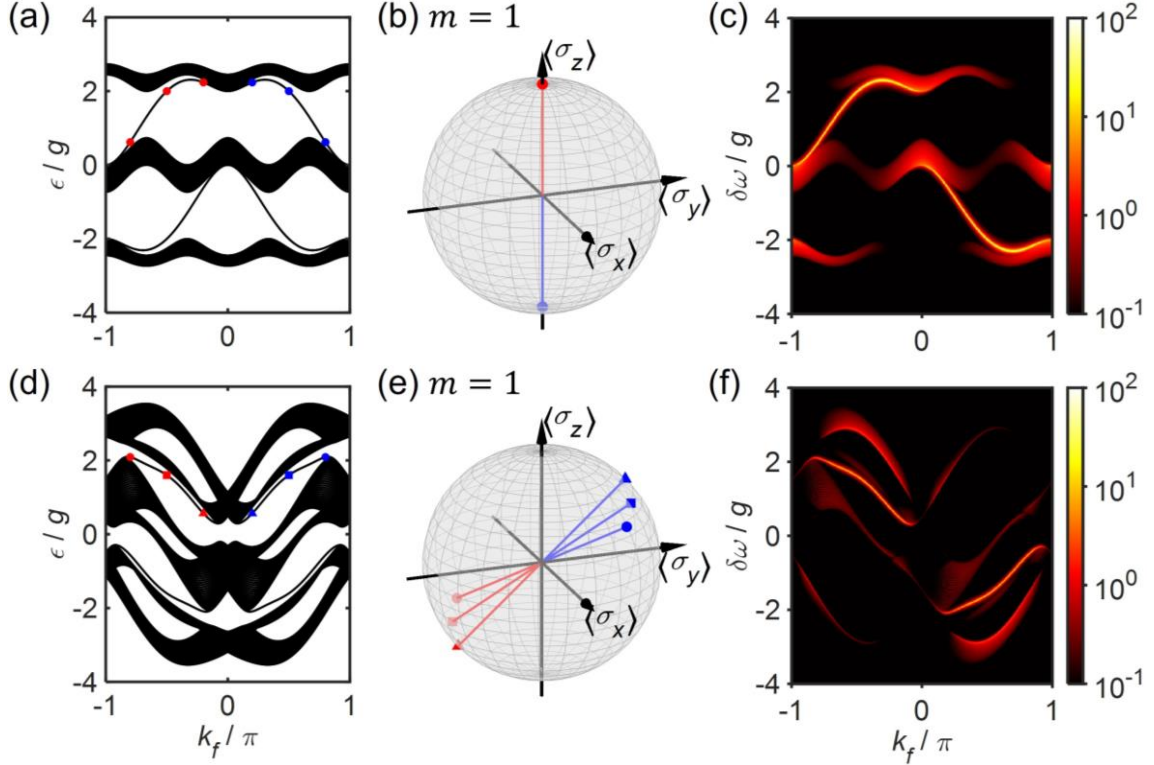

**Fig. 3.** Eigenspectra of the Hamiltonian $\hat{H}_1$ under the open boundary condition. (a)–(c) The Abelian case, $(\theta, \varphi) = (0, 2\pi/3)$. (d)–(f) The non-Abelian case, $(\theta, \varphi) = (2\pi/3, 2\pi/3)$. (a) and (d) are eigenspectra by diagonalizing $\hat{H}_1$. (b) and (e) are pseudo-spins of the edge states on the 1st lattice site, with $k_f$ and $\epsilon$ values indicated by the red and blue dots in (a) and (d). (c) and (f) are numerically simulated results of $N(k_f, \delta\omega)$ with $\gamma = \gamma_0 = 0.01g$ and $m_0 = 1$. $\kappa = g$, $M = 30q - 1 = 89$ in this figure.

Finally, we show in Fig. 4 that by performing time-averaged measurements, we can resolve the Hofstadter butterfly spectrum associated with the Hamiltonian $\hat{H}_1$. Given the parameter combination $(\theta, \varphi)$ in the Hamiltonian, we measure the photon number in all cavities averaged within a round-trip time:

$$\bar{N}(\delta\omega; \theta, \varphi) \equiv \frac{1}{2\pi} \int_{-\pi}^{\pi} N(k_f, \delta\omega) \mathrm{d}k_f$$

(17)

Eq. (17) is a time-averaged measurement since $k_f$ is interpreted as the fast time variable $t$. For any $(\theta, \varphi)$ value, the Hamiltonian $\hat{H}_1(\theta, \varphi)$ can be implemented by our synthetic frequency dimension platform via the mathematical mapping in Eq. (8). Such flexibility enables us to probe $\bar{N}(\delta\omega; \theta, \varphi)$ in the full parameter space $0 \leq \theta, \varphi < 2\pi$. Figure 4 shows two representative slices of the parameter space, $\theta = 0$ [Abelian case, Figs. 4(a) and 4(b)] and $\theta = 2\pi/3$ [non-Abelian case, Figs. 4(c) and 4(d)]. In Figs. 4(b) and 4(d), we plot the numerically simulated results of $\bar{N}(\delta\omega; \theta, \varphi)$. The resonance locations of $\bar{N}(\delta\omega; \theta, \varphi)$ correspond to the allowed energies in Figs. 4(a) and 4(c) [89].

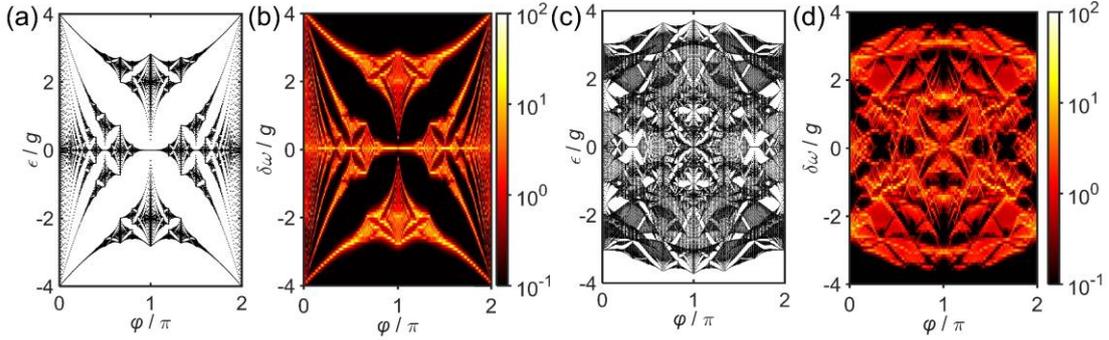

**Fig. 4.** Hofstadter butterfly spectra of the Hamiltonian $\hat{H}_1$. (a), (b) The Abelian case, $\theta = 0$. (c), (d) The non-Abelian case, $\theta = 2\pi/3$. (a) and (c) are acquired by diagonalizing the Hamiltonian $\hat{H}_1$. (b) and (d) are simulated results of $\bar{N}(\delta\omega; \theta, \varphi)$. $\kappa = g$, $M = 255$ in this figure, $\gamma = \gamma_0 = 0.01g$, $m_0 = 1$, $\psi_\uparrow^{\text{exc}} = 1$, $\psi_\downarrow^{\text{exc}} = 0$ in (b) and (d).

*Conclusion*—We have developed a scheme to artificially synthesize arbitrary SU(2) lattice gauge fields for photons, using the concept of synthetic frequency dimensions in a coupled resonator array. The parameters of the non-Abelian gauge field are flexible by tuning the modulation parameters and polarization rotation parameters in our optical setup. We also numerically demonstrate that the band structures of the lattice can be resolved by steady-state measurements. Such eigenspectra exhibit distinct features when the underlying lattice gauge field becomes non-Abelian. The results presented in this Letter provide opportunities to explore novel topological phenomena associated with non-Abelian lattice gauge fields in photonic systems.

**Acknowledgements**

The authors are grateful to Prof. Avik Dutt, Prof. Yi Yang and Prof. Luqi Yuan for helpful discussions. This work is supported by MURI projects from the US Air Force Office of Scientific Research (grant no. FA9550-18-1-0379, FA FA9550-22-1-0339), and by a Vannevar Bush Faculty Fellowship from the US Department of Defense (grant no. N00014-17-1-3030).

**References**

[1] K. v. Klitzing, G. Dorda, and M. Pepper, *New Method for High-Accuracy Determination of the Fine-Structure Constant Based on Quantized Hall Resistance*, Physical Review Letters **45**, 494 (1980).

[2] C. L. Kane and E. J. Mele, *Quantum Spin Hall Effect in Graphene*, Physical Review Letters **95**, 226801 (2005).

[3] B. A. Bernevig and S.-C. Zhang, *Quantum Spin Hall Effect*, Physical Review Letters **96**, 106802 (2006).

[4] N. Goldman, A. Kubasiak, A. Bermudez, P. Gaspard, M. Lewenstein, and M. A. Martin-Delgado, *Non-Abelian Optical Lattices: Anomalous Quantum Hall Effect and Dirac Fermions*, Physical Review Letters **103**, 035301 (2009).

[5] P. Hauke, O. Tieleman, A. Celi, C. Ölschläger, J. Simonet, J. Struck, M. Weinberg, P. Windpassinger, K. Sengstock, M. Lewenstein *et al.*, *Non-Abelian Gauge Fields and Topological Insulators in Shaken Optical Lattices*, Physical Review Letters **109**, 145301 (2012).

[6] T. Iadecola, T. Schuster, and C. Chamon, *Non-Abelian Braiding of Light*, Physical Review Letters **117**, 073901 (2016).

[7] S. Sugawa, F. Salces-Carcoba, R. Perry Abigail, Y. Yue, and I. B. Spielman, *Second Chern number of a quantum-simulated non-Abelian Yang monopole*, Science **360**, 1429 (2018).

[8] E. Yang, B. Yang, O. You, H.-C. Chan, P. Mao, Q. Guo, S. Ma, L. Xia, D. Fan, Y. Xiang *et al.*, *Observation of Non-Abelian Nodal Links in Photonics*, Physical Review Letters **125**, 033901 (2020).

[9] Y. Yang, B. Zhen, J. D. Joannopoulos, and M. Soljačić, *Non-Abelian generalizations of the Hofstadter model: spin–orbit-coupled butterfly pairs*, Light: Science & Applications **9**, 177 (2020).

[10] M. Di Liberto, N. Goldman, and G. Palumbo, *Non-Abelian Bloch oscillations in higher-order topological insulators*, Nature Communications **11**, 5942 (2020).

[11] Q. Guo, T. Jiang, R.-Y. Zhang, L. Zhang, Z.-Q. Zhang, B. Yang, S. Zhang, and C. T. Chan, *Experimental observation of non-Abelian topological charges and edge states*, Nature **594**, 195 (2021).

[12] T. Jiang, Q. Guo, R.-Y. Zhang, Z.-Q. Zhang, B. Yang, and C. T. Chan, *Four-band non-Abelian topological insulator and its experimental realization*, Nature Communications **12**, 6471 (2021).

[13] Z.-G. Chen, R.-Y. Zhang, C. T. Chan, and G. Ma, *Classical non-Abelian braiding of acoustic modes*, Nature Physics **18**, 179 (2022).

[14] X.-L. Zhang, F. Yu, Z.-G. Chen, Z.-N. Tian, Q.-D. Chen, H.-B. Sun, and G. Ma, *Non-Abelian braiding on photonic chips*, Nature Photonics **16**, 390 (2022).

[15] M. Onoda, S. Murakami, and N. Nagaosa, *Geometrical aspects in optical wave-packet dynamics*, Physical Review E **74**, 066610 (2006).

[16] K. Y. Bliokh, D. Y. Frolov, and Y. A. Kravtsov, *Non-Abelian evolution of electromagnetic waves in a weakly anisotropic inhomogeneous medium*, Physical Review A **75**, 053821 (2007).

[17] R. O. Umucalılar and I. Carusotto, *Artificial gauge field for photons in coupled cavity arrays*, Physical Review A **84**, 043804 (2011).

[18] K. Fang, Z. Yu, and S. Fan, *Realizing effective magnetic field for photons by controlling the phase of dynamic modulation*, Nature Photonics **6**, 782 (2012).

[19] L. D. Tzuang, K. Fang, P. Nussenzveig, S. Fan, and M. Lipson, *Non-reciprocal phase shift induced by an effective magnetic flux for light*, Nature Photonics **8**, 701 (2014).

[20] L. Yuan, Y. Shi, and S. Fan, *Photonic gauge potential in a system with a synthetic frequency dimension*, Opt. Lett. **41**, 741 (2016).

[21] Y. Chen, R.-Y. Zhang, Z. Xiong, Z. H. Hang, J. Li, J. Q. Shen, and C. T. Chan, *Non-Abelian gauge field optics*, Nature Communications **10**, 3125 (2019).

[22] Y. Yang, C. Peng, D. Zhu, H. Buljan, J. D. Joannopoulos, B. Zhen, and M. Soljačić, *Synthesis and observation of non-Abelian gauge fields in real space*, Science **365**, 1021 (2019).

[23] V. Brosco, L. Pilozzi, R. Fazio, and C. Conti, *Non-Abelian Thouless pumping in a photonic lattice*, Physical Review A **103**, 063518 (2021).

[24] K. Y. Bliokh and V. D. Freilikher, *Polarization transport of transverse acoustic waves: Berry phase and spin Hall effect of phonons*, Physical Review B **74**, 174302 (2006).

[25] M. Nash Lisa, D. Kleckner, A. Read, V. Vitelli, M. Turner Ari, and T. M. Irvine William, *Topological mechanics of gyroscopic metamaterials*, Proceedings of the National Academy of Sciences **112**, 14495 (2015).

[26] M. Xiao, W.-J. Chen, W.-Y. He, and C. T. Chan, *Synthetic gauge flux and Weyl points in acoustic systems*, Nature Physics **11**, 920 (2015).

[27] G. Ma, M. Xiao, and C. T. Chan, *Topological phases in acoustic and mechanical systems*, Nature Reviews Physics **1**, 281 (2019).

[28] M. Fruchart, Y. Zhou, and V. Vitelli, *Dualities and non-Abelian mechanics*, Nature **577**, 636 (2020).

[29] J. P. Mathew, J. d. Pino, and E. Verhagen, *Synthetic gauge fields for phonon transport in a nano-optomechanical system*, Nature Nanotechnology **15**, 198 (2020).

[30] Y. Chen, Y.-L. Zhang, Z. Shen, C.-L. Zou, G.-C. Guo, and C.-H. Dong, *Synthetic Gauge Fields in a Single Optomechanical Resonator*, Physical Review Letters **126**, 123603 (2021).

[31] K. Osterloh, M. Baig, L. Santos, P. Zoller, and M. Lewenstein, *Cold Atoms in Non-Abelian Gauge Potentials: From the Hofstadter "Moth" to Lattice Gauge Theory*, Physical Review Letters **95**, 010403 (2005).

[32]     J. Ruseckas, G. Juzeliūnas, P. Öhberg, and M. Fleischhauer, *Non-Abelian Gauge Potentials for Ultracold Atoms with Degenerate Dark States*, Physical Review Letters **95**, 010404 (2005).

[33]     G. Juzeliūnas, J. Ruseckas, A. Jacob, L. Santos, and P. Öhberg, *Double and Negative Reflection of Cold Atoms in Non-Abelian Gauge Potentials*, Physical Review Letters **100**, 200405 (2008).

[34]     N. Goldman, A. Kubasiak, P. Gaspard, and M. Lewenstein, *Ultracold atomic gases in non-Abelian gauge potentials: The case of constant Wilson loop*, Physical Review A **79**, 023624 (2009).

[35]     Y. J. Lin, R. L. Compton, K. Jiménez-García, J. V. Porto, and I. B. Spielman, *Synthetic magnetic fields for ultracold neutral atoms*, Nature **462**, 628 (2009).

[36]     J. Dalibard, F. Gerbier, G. Juzeliūnas, and P. Öhberg, *Colloquium: Artificial gauge potentials for neutral atoms*, Reviews of Modern Physics **83**, 1523 (2011).

[37]     M. Aidelsburger, M. Atala, S. Nascimbène, S. Trotzky, Y. A. Chen, and I. Bloch, *Experimental Realization of Strong Effective Magnetic Fields in an Optical Lattice*, Physical Review Letters **107**, 255301 (2011).

[38]     J. Struck, C. Ölschläger, M. Weinberg, P. Hauke, J. Simonet, A. Eckardt, M. Lewenstein, K. Sengstock, and P. Windpassinger, *Tunable Gauge Potential for Neutral and Spinless Particles in Driven Optical Lattices*, Physical Review Letters **108**, 225304 (2012).

[39]     H. Miyake, G. A. Siviloglou, C. J. Kennedy, W. C. Burton, and W. Ketterle, *Realizing the Harper Hamiltonian with Laser-Assisted Tunneling in Optical Lattices*, Physical Review Letters **111**, 185302 (2013).

[40]     A. Kosior and K. Sacha, *Simulation of non-Abelian lattice gauge fields with a single-component gas*, EPL (Europhysics Letters) **107**, 26006 (2014).

[41]     N. Goldman, G. Juzeliūnas, P. Öhberg, and I. B. Spielman, *Light-induced gauge fields for ultracold atoms*, Reports on Progress in Physics **77**, 126401 (2014).

[42]     E. G. Guan, H. Yu, and G. Wang, *Non-Abelian gauge potentials driven localization transition in quasiperiodic optical lattices*, Physics Letters A **384**, 126152 (2020).

[43]     S. Sugawa, F. Salces-Carcoba, Y. Yue, A. Putra, and I. B. Spielman, *Wilson loop and Wilczek-Zee phase from a non-Abelian gauge field*, npj Quantum Information **7**, 144 (2021).

[44]     M. Aidelsburger, S. Nascimbene, and N. Goldman, *Artificial gauge fields in materials and engineered systems*, Comptes Rendus Physique **19**, 394 (2018).

[45]     H. Terças, H. Flayac, D. D. Solnyshkov, and G. Malpuech, *Non-Abelian Gauge Fields in Photonic Cavities and Photonic Superfluids*, Physical Review Letters **112**, 066402 (2014).

[46]     D. Solnyshkov, A. Nalitov, B. Teklu, L. Franck, and G. Malpuech, *Spin-dependent Klein tunneling in polariton graphene with photonic spin-orbit interaction*, Physical Review B **93**, 085404 (2016).

[47] L. B. Ma, S. L. Li, V. M. Fomin, M. Hentschel, J. B. Götte, Y. Yin, M. R. Jorgensen, and O. G. Schmidt, *Spin–orbit coupling of light in asymmetric microcavities*, Nature Communications **7**, 10983 (2016).

[48] A. Gianfrate, O. Bleu, L. Dominici, V. Ardizzone, M. De Giorgi, D. Ballarini, G. Lerario, K. W. West, L. N. Pfeiffer, D. D. Solnyshkov *et al.*, *Measurement of the quantum geometric tensor and of the anomalous Hall drift*, Nature **578**, 381 (2020).

[49] C. E. Whittaker, T. Dowling, A. V. Nalitov, A. V. Yulin, B. Royall, E. Clarke, M. S. Skolnick, I. A. Shelykh, and D. N. Krizhanovskii, *Optical analogue of Dresselhaus spin–orbit interaction in photonic graphene*, Nature Photonics **15**, 193 (2021).

[50] L. Polimeno, A. Fieramosca, G. Lerario, L. De Marco, M. De Giorgi, D. Ballarini, L. Dominici, V. Ardizzone, M. Pugliese, C. T. Prontera *et al.*, *Experimental investigation of a non-Abelian gauge field in 2D perovskite photonic platform*, Optica **8**, 1442 (2021).

[51] S. Spencer Michael, Y. Fu, P. Schlaus Andrew, D. Hwang, Y. Dai, D. Smith Matthew, R. Gamelin Daniel, and X. Y. Zhu, *Spin-orbit–coupled exciton-polariton condensates in lead halide perovskites*, Science Advances **7**, eabj7667 (2021).

[52] A. A. Abdumalikov Jr, J. M. Fink, K. Juliusson, M. Pechal, S. Berger, A. Wallraff, and S. Filipp, *Experimental realization of non-Abelian non-adiabatic geometric gates*, Nature **496**, 482 (2013).

[53] Z. Song, T. Wu, W. Wu, and R. Yu, *Experimental realization of non-Abelian gauge potentials and topological Chern state in circuit system*, arXiv:2009.04870 [cond-mat.mes-hall] (2020).

[54] F. de Juan, *Non-Abelian gauge fields and quadratic band touching in molecular graphene*, Physical Review B **87**, 125419 (2013).

[55] O. Boada, A. Celi, J. I. Latorre, and M. Lewenstein, *Quantum Simulation of an Extra Dimension*, Physical Review Letters **108**, 133001 (2012).

[56] A. Celi, P. Massignan, J. Ruseckas, N. Goldman, I. B. Spielman, G. Juzeliūnas, and M. Lewenstein, *Synthetic Gauge Fields in Synthetic Dimensions*, Physical Review Letters **112**, 043001 (2014).

[57] M. Mancini, G. Pagano, G. Cappellini, L. Livi, M. Rider, J. Catani, C. Sias, P. Zoller, M. Inguscio, M. Dalmonte *et al.*, *Observation of chiral edge states with neutral fermions in synthetic Hall ribbons*, Science **349**, 1510 (2015).

[58] T. Ozawa, H. M. Price, N. Goldman, O. Zilberberg, and I. Carusotto, *Synthetic dimensions in integrated photonics: From optical isolation to four-dimensional quantum Hall physics*, Physical Review A **93**, 043827 (2016).

[59] L. Yuan and S. Fan, *Bloch oscillation and unidirectional translation of frequency in a dynamically modulated ring resonator*, Optica **3**, 1014 (2016).

[60] X.-W. Luo, X. Zhou, J.-S. Xu, C.-F. Li, G.-C. Guo, C. Zhang, and Z.-W. Zhou, *Synthetic-lattice enabled all-optical devices based on orbital angular momentum of light*, Nature Communications **8**, 16097 (2017).

[61] B. A. Bell, K. Wang, A. S. Solntsev, D. N. Neshev, A. A. Sukhorukov, and B. J. Eggleton, *Spectral photonic lattices with complex long-range coupling*, Optica **4**, 1433 (2017).

[62] O. Zilberberg, S. Huang, J. Guglielmon, M. Wang, K. P. Chen, Y. E. Kraus, and M. C. Rechtsman, *Photonic topological boundary pumping as a probe of 4D quantum Hall physics*, Nature **553**, 59 (2018).

[63] L. Yuan, M. Xiao, Q. Lin, and S. Fan, *Synthetic space with arbitrary dimensions in a few rings undergoing dynamic modulation*, Physical Review B **97**, 104105 (2018).

[64] C. Qin, F. Zhou, Y. Peng, D. Sounas, X. Zhu, B. Wang, J. Dong, X. Zhang, A. Alù, and P. Lu, *Spectrum Control through Discrete Frequency Diffraction in the Presence of Photonic Gauge Potentials*, Physical Review Letters **120**, 133901 (2018).

[65] L. Yuan, Q. Lin, M. Xiao, A. Dutt, and S. Fan, *Pulse shortening in an actively mode-locked laser with parity-time symmetry*, APL Photonics **3**, 086103 (2018).

[66] L. Yuan, Q. Lin, M. Xiao, and S. Fan, *Synthetic dimension in photonics*, Optica **5**, 1396 (2018).

[67] E. Lustig, S. Weimann, Y. Plotnik, Y. Lumer, M. A. Bandres, A. Szameit, and M. Segev, *Photonic topological insulator in synthetic dimensions*, Nature **567**, 356 (2019).

[68] T. Ozawa and H. M. Price, *Topological quantum matter in synthetic dimensions*, Nature Reviews Physics **1**, 349 (2019).

[69] H. Chalabi, S. Barik, S. Mittal, T. E. Murphy, M. Hafezi, and E. Waks, *Synthetic Gauge Field for Two-Dimensional Time-Multiplexed Quantum Random Walks*, Physical Review Letters **123**, 150503 (2019).

[70] A. Dutt, Q. Lin, L. Yuan, M. Minkov, M. Xiao, and S. Fan, *A single photonic cavity with two independent physical synthetic dimensions*, Science **367**, 59 (2020).

[71] C. Joshi, A. Farsi, A. Dutt, B. Y. Kim, X. Ji, Y. Zhao, A. M. Bishop, M. Lipson, and A. L. Gaeta, *Frequency-Domain Quantum Interference with Correlated Photons from an Integrated Microresonator*, Physical Review Letters **124**, 143601 (2020).

[72] H. Chalabi, S. Barik, S. Mittal, T. E. Murphy, M. Hafezi, and E. Waks, *Guiding and confining of light in a two-dimensional synthetic space using electric fields*, Optica **7**, 506 (2020).

[73] A. Dutt, M. Minkov, I. A. D. Williamson, and S. Fan, *Higher-order topological insulators in synthetic dimensions*, Light: Science & Applications **9**, 131 (2020).

[74] K. Wang, B. A. Bell, A. S. Solntsev, D. N. Neshev, B. J. Eggleton, and A. A. Sukhorukov, *Multidimensional synthetic chiral-tube lattices via nonlinear frequency conversion*, Light: Science & Applications **9**, 132 (2020).

[75] Y. Hu, C. Reimer, A. Shams-Ansari, M. Zhang, and M. Loncar, *Realization of high-dimensional frequency crystals in electro-optic microcombs*, Optica **7**, 1189 (2020).

[76] K. Wang, A. Dutt, K. Y. Yang, C. C. Wojcik, J. Vučković, and S. Fan, *Generating arbitrary topological windings of a non-Hermitian band*, Science **371**, 1240 (2021).

[77] E. Lustig and M. Segev, *Topological photonics in synthetic dimensions*, Advances in Optics and Photonics **13**, 426 (2021).

[78] D. Cheng, B. Peng, D.-W. Wang, X. Chen, L. Yuan, and S. Fan, *Arbitrary synthetic dimensions via multiboson dynamics on a one-dimensional lattice*, Physical Review Research **3**, 033069 (2021).

[79] K. Wang, A. Dutt, C. C. Wojcik, and S. Fan, *Topological complex-energy braiding of non-Hermitian bands*, Nature **598**, 59 (2021).

[80] C. Leefmans, A. Dutt, J. Williams, L. Yuan, M. Parto, F. Nori, S. Fan, and A. Marandi, *Topological dissipation in a time-multiplexed photonic resonator network*, Nature Physics **18**, 442 (2022).

[81] R. Simon and N. Mukunda, *Minimal three-component SU(2) gadget for polarization optics*, Physics Letters A **143**, 165 (1990).

[82] A. Dutt, M. Minkov, Q. Lin, L. Yuan, D. A. B. Miller, and S. Fan, *Experimental band structure spectroscopy along a synthetic dimension*, Nature Communications **10**, 3122 (2019).

[83] K. Wang, A. Dutt, Y. Yang Ki, C. Wojcik Casey, J. Vučković, and S. Fan, *Generating arbitrary topological windings of a non-Hermitian band*, Science **371**, 1240 (2021).

[84] G. Li, Y. Zheng, A. Dutt, D. Yu, Q. Shan, S. Liu, L. Yuan, S. Fan, and X. Chen, *Dynamic band structure measurement in the synthetic space*, Science Advances **7**, eabe4335 (2021).

[85] L. Yuan, A. Dutt, and S. Fan, *Synthetic frequency dimensions in dynamically modulated ring resonators*, APL Photonics **6**, 071102 (2021).

[86] A. Dutt, L. Yuan, K. Y. Yang, K. Wang, S. Buddhiraju, J. Vučković, and S. Fan, *Creating boundaries along a synthetic frequency dimension*, Nature Communications **13**, 3377 (2022).

[87] C. W. Gardiner and M. J. Collett, *Input and output in damped quantum systems: Quantum stochastic differential equations and the master equation*, Physical Review A **31**, 3761 (1985).

[88] S. Fan, Ş. E. Kocabaş, and J.-T. Shen, *Input-output formalism for few-photon transport in one-dimensional nanophotonic waveguides coupled to a qubit*, Physical Review A **82**, 063821 (2010).

[89] See Supplemental Materials for more discussions.

[90] M. Z. Hasan and C. L. Kane, *Colloquium: Topological insulators*, Reviews of Modern Physics **82**, 3045 (2010).

**Supplemental Material of**

**Artificial non-Abelian lattice gauge fields for photons in the synthetic frequency dimension**

**Discussions on the band structures in Fig. 2**

As shown by Fig. 2(a), in the Abelian case, each of the three bands of Hamiltonian $\hat{H}_1$ is two-fold degenerate. Such degeneracy is protected by an anti-unitary symmetry operator $\mathcal{U}$ that satisfies $\mathcal{U}^2 = -1$:

$$\mathcal{U}\hat{H}_1^{\text{PBC}}(k_m, k_f)\mathcal{U}^{-1} = \hat{H}_1^{\text{PBC}}(k_m, k_f), \qquad \mathcal{U} = i\sigma_y \mathcal{K} \otimes \begin{bmatrix} & & 1 \\ & \ddots & \\ 1 & & \\ & & 1 \end{bmatrix} \tag{S18}$$

where $\mathcal{K}$ is the complex conjugate operator. The superscript PBC indicates the periodic boundary condition. In the non-Abelian case, however, the mixture of the spin-up and spin-down states breaks such symmetry except at $k_f = 0, \pi$. Therefore, each degenerate band in the Abelian case is split into two bands in the non-Abelian case, as colored in red and blue in Fig. 2(e), respectively.

**Discussions on the eigenspectra in Fig. 3**

In Figs. 3(a) and 3(d), the edge states are approximately two-fold degenerate for both the Abelian and the non-Abelian cases. In contrast, the bulk states are non-degenerate in the non-Abelian case. The approximate degeneracy of the edge states arises because the system has the spatial inversion symmetry

$$\Sigma_x \hat{H}_1^{\text{OBC}}(k_f) \Sigma_x^{-1} = \hat{H}_1^{\text{OBC}}(k_f), \qquad \Sigma_x = \sigma_x \otimes \begin{bmatrix} & & 1 \\ & \ddots & \\ 1 & & \end{bmatrix} \tag{S19}$$

The superscript OBC indicates the open boundary condition. A state localized on one edge of the lattice with a certain spin is mapped by the $\Sigma_x$ operation to the state localized on the other edge with the opposite spin. Such degeneracy is dependent on the truncation. In the calculations of Fig. 3, the lattice includes $M = 30q - 1 = 89$ sites along the spatial dimension. In general, $M \equiv -1 \pmod{q}$ is necessary for Eq. (S19). For other site numbers $M$, the spatial inversion symmetry represented by $\Sigma_x$ is broken, and the approximate degeneracy of the edge states is lifted.

Regarding the spin-dependency of the edge states of $\hat{H}_1^{\text{OBC}}$ in the non-Abelian case, here we examine three related properties. They are illustrated in the Supplementary Video and supported by arguments based on symmetry analysis. We consider an edge state located at $(k_f, \epsilon)$, and write it as $|\psi^{\text{eig}}(k_f)\rangle \equiv \left[\psi_\uparrow^{\text{eig}}(1, k_f, \epsilon), \psi_\downarrow^{\text{eig}}(1, k_f, \epsilon), \cdots, \psi_\uparrow^{\text{eig}}(M, k_f, \epsilon), \psi_\downarrow^{\text{eig}}(M, k_f, \epsilon)\right]^{\text{T}}$. We also define the pseudo-spin state on the $m$-th lattice site as $|\psi^{\text{eig}}(m, k_f)\rangle \equiv \left[\psi_\uparrow^{\text{eig}}(m, k_f, \epsilon), \psi_\downarrow^{\text{eig}}(m, k_f, \epsilon)\right]^{\text{T}}$. Associated with this particular edge state, we observe the following properties:

(1)    there is another edge state at $(-k_f, \epsilon)$ with orthogonal pseudo-spin states on each lattice site, i.e., $\langle\psi^{\text{eig}}(m, -k_f)|\psi^{\text{eig}}(m, k_f)\rangle = 0$;

(2)    there is another edge state at $(k_f + \pi, -\epsilon)$, and their pseudo-spin states on each lattice site are at the same point on the Bloch sphere;

(3)    there is another edge state at $(\pi - k_f, -\epsilon)$ with orthogonal pseudo-spin states on each lattice site, i.e., $\langle\psi^{\text{eig}}(m, \pi - k_f)|\psi^{\text{eig}}(m, k_f)\rangle = 0$.

The Property (1) is the result of Kramers' degeneracy. Indeed, the system has the symmetry

$$\hat{H}_1^{\text{OBC}}(-k_f) = (i\sigma_y \mathcal{K} \otimes \mathbb{I}_M)\hat{H}_1^{\text{OBC}}(k_f)(i\sigma_y \mathcal{K} \otimes \mathbb{I}_M)^{-1}$$

(S20)

where $\mathcal{K}$ is the complex conjugate operation. Eq. (S3) describes a local symmetry, so on each lattice site we have

$$|\psi^{\text{eig}}(m, -k_f)\rangle = i\sigma_y \mathcal{K} |\psi^{\text{eig}}(m, k_f)\rangle$$

(S21)

Since $(i\sigma_y \mathcal{K})^2 = -\sigma_0$, we know from the Kramers' Theorem that $\langle\psi^{\text{eig}}(m, -k_f)|\psi^{\text{eig}}(m, k_f)\rangle = 0$. The Property (2) is the result of the chiral symmetry:

$$\hat{H}_1^{\text{OBC}}(k_f + \pi) = -\mathcal{C}\hat{H}_1^{\text{OBC}}(k_f)\mathcal{C}^{-1}$$

(S22)

where $\mathcal{C}$ is a diagonal matrix with $(-1)^m$ as the $m$-th element along the diagonal. Again, Eq. (S5) describes a local symmetry, and the eigenstate $|\psi^{\text{eig}}(m, k_f + \pi)\rangle$ is then related to $|\psi^{\text{eig}}(m, k_f)\rangle$ by

$$|\psi^{\text{eig}}(m, k_f + \pi)\rangle = (-1)^m |\psi^{\text{eig}}(m, k_f)\rangle$$

(S23)

which is a spin-independent local gauge transform. Therefore, $|\psi^{\text{eig}}(m, k_f)\rangle$ and $|\psi^{\text{eig}}(m, k_f + \pi)\rangle$ are at the same point on the Bloch sphere. And Property (3) can be deduced from Properties (1) and (2).

In the Supplementary Video, we plot, on the Bloch sphere, the pseudo-spins of the edge states on the lattice site numbers 1–4 located at $(k_f, \epsilon)$, $(-k_f, \epsilon)$, $(k_f + \pi, -\epsilon)$ and $(\pi - k_f, -\epsilon)$. The video shows their orthogonality, which explains the chiral excitations in Figs. 3(c) and 3(f) in the main text.

**Discussions on the butterfly spectra in Fig. 4**

In the Abelian case, the allowed energies with respect to the parameter $\varphi$ form the standard Hofstadter butterfly as shown in Fig. 4(a). This spectrum includes two degenerate copies of butterflies associated with spin-up and spin-down states. In the non-Abelian case of Fig. 4(c), the allowed energy spectrum becomes more complicated. Repetitions of the butterfly structures at different scales are present, and the forbidden energies form holes of various sizes in between. Similar observations have been made in the 'Hofstadter moth spectrum' [1].

**Reference**

[1]     K. Osterloh, M. Baig, L. Santos, P. Zoller, and M. Lewenstein, *Cold Atoms in Non-Abelian Gauge Potentials: From the Hofstadter "Moth" to Lattice Gauge Theory*, Physical Review Letters **95**, 010403 (2005).